\documentclass[sigconf]{acmart}
 \usepackage{amsmath,amssymb,amsfonts,bm}
\usepackage{graphicx}
\usepackage{textcomp}
\usepackage{xcolor}
\usepackage{float}
\usepackage{array}
\usepackage{tabularx}
\usepackage{array}
\usepackage{balance}
\usepackage{marvosym}
\usepackage{multirow}
\usepackage{array}
\usepackage{enumitem}
\usepackage{booktabs}
\usepackage{algpseudocode}
\usepackage[ruled]{algorithm2e} 
\usepackage{xcolor}
\usepackage{pifont}
\usepackage{url}
\usepackage{subfig}
\usepackage{makecell}
\DeclareMathOperator{\sign}{sign}

\AtBeginDocument{%
  \providecommand\BibTeX{{%
    \normalfont B\kern-0.5em{\scshape i\kern-0.25em b}\kern-0.8em\TeX}}}

\copyrightyear{2022} 
\acmYear{2022} 
\setcopyright{acmcopyright}
\acmConference[SIGIR '22]{Proceedings of the 45th International ACM SIGIR Conference on Research and Development in Information Retrieval}{July 11--15, 2022}{Madrid, Spain}
\acmBooktitle{Proceedings of the 45th International ACM SIGIR Conference on Research and Development in Information Retrieval (SIGIR '22), July 11--15, 2022, Madrid, Spain}
\acmPrice{15.00}
\acmDOI{10.1145/3477495.3531937}
\acmISBN{978-1-4503-8732-3/22/07}

\begin{document}
\fancyhead{}

\title{Are Graph Augmentations Necessary? Simple Graph Contrastive Learning for Recommendation}


\author{Junliang Yu}
\affiliation{%
	\institution{The University of Queensland}
	\city{Brisbane}
	\country{Australia}}
\email{jl.yu@uq.edu.au}

\author{Hongzhi Yin}
\authornote{Corresponding author.}
\affiliation{%
	\institution{The University of Queensland}
	\city{Brisbane}
	\country{Australia}}
\email{h.yin1@uq.edu.au}

\author{Xin Xia}
\affiliation{%
	\institution{The University of Queensland}
	\city{Brisbane}
	\country{Australia}}
\email{x.xia@uq.edu.au}

\author{Tong Chen}
\affiliation{%
	\institution{The University of Queensland}
	\city{Brisbane}
	\country{Australia}}
\email{tong.chen@uq.edu.au}

\author{Lizhen Cui}
\affiliation{%
	\institution{Shandong University}
	\city{Jinan}
	\country{China}}
\email{clz@sdu.edu.cn}

\author{Nguyen Quoc Viet Hung}
\affiliation{%
	\institution{Griffith University}
		\city{Gold Coast}
	\country{Australia}}
\email{quocviethung1@gmail.com}

\begin{abstract}
	Contrastive learning (CL) recently has spurred a fruitful line of research in the field of recommendation, since its ability to extract self-supervised signals from the raw data is well-aligned with recommender systems' needs for tackling the data sparsity issue. A typical pipeline of CL-based recommendation models is first augmenting the user-item bipartite graph with structure perturbations, and then maximizing the node representation consistency between different graph augmentations. Although this paradigm turns out to be effective, what underlies the performance gains is still a mystery. In this paper, we first experimentally disclose that, in CL-based recommendation models, CL operates by learning more evenly distributed user/item representations that can implicitly mitigate the popularity bias. Meanwhile, we reveal that the graph augmentations, which were considered necessary, just play a trivial role. Based on this finding, we propose a simple CL method which discards the graph augmentations and instead adds uniform noises to the embedding space for creating contrastive views. A comprehensive experimental study on three benchmark datasets demonstrates that, though it appears strikingly simple, the proposed method can smoothly adjust the uniformity of learned representations and has distinct advantages over its graph augmentation-based counterparts in terms of recommendation accuracy and training efficiency. The code is released at \url{https://github.com/Coder-Yu/QRec}.
\end{abstract}


\maketitle

\begin{figure}[t]
	\centering
	\includegraphics[width=.45\textwidth]{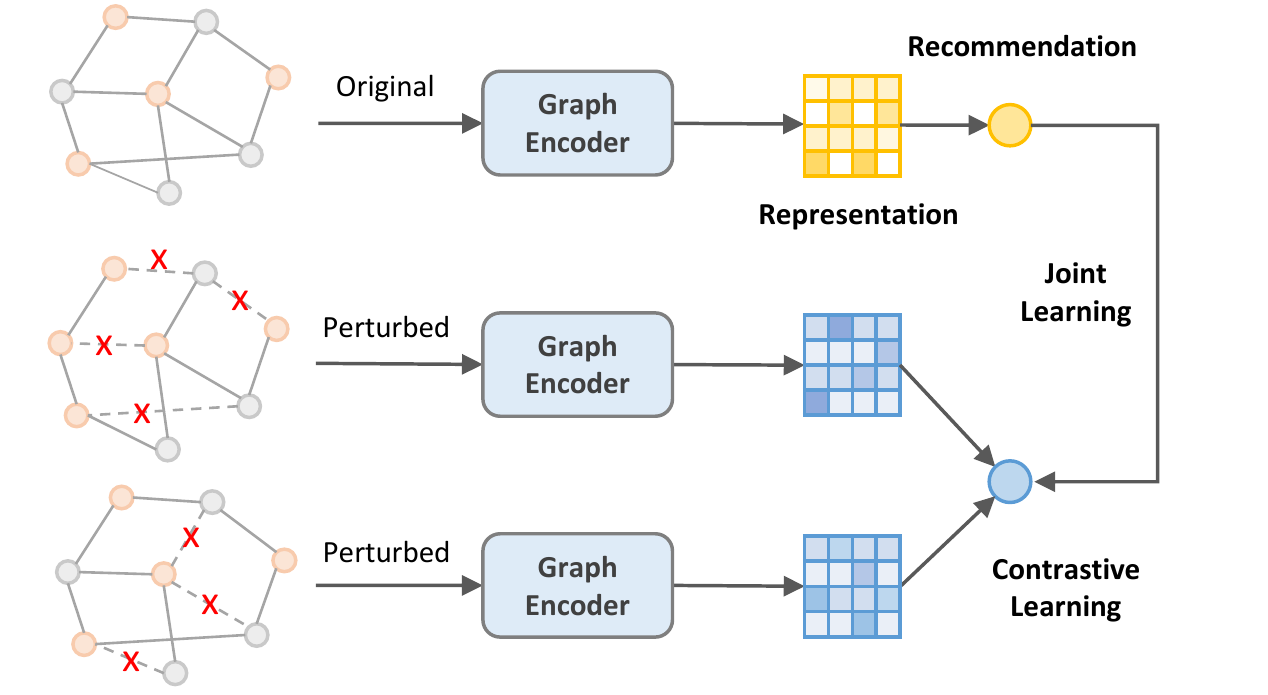}
	\caption{Graph contrastive learning with edge dropout for recommendation. }
	\label{figure:gcl}
	\vspace{-15px}		
\end{figure}

\section{Introduction}
Recently, a resurgence of contrastive learning (CL) \cite{DBLP:conf/nips/KhoslaTWSTIMLK20,jaiswal2021survey,liu2020self} has been witnessed in deep representation learning. Due to the ability to extract the general features from massive unlabeled data and regularize representations in a self-supervised manner, CL has led to major advances in multiple research fields \cite{you2020graph,wujc2021self,chen2020simple,gao2021simcse}. As data annotation is not required in CL, it is a natural antidote to the data sparsity issue in recommender systems \cite{sarwar2001sparsity,hao2021pre}. An increasing number of very recent studies \cite{wujc2021self,yu2021self,xia2020self,zhou2020s,zhou2021contrastive,yu2021socially} have sought to harness CL for improving recommendation performance and have demonstrated significant gains. A typical way \cite{wujc2021self} to apply CL to recommendation is first augmenting the user-item bipartite graph with structure perturbations (e.g., stochastic edge/node dropout at a certain ratio), and then maximizing the consistency of representations under different views learned via a graph encoder. In this setting, the CL task acts as the auxiliary task, and is jointly optimized with the recommendation task (Fig. \ref{figure:gcl}). \par

Despite the encouraging results achieved by CL, however, what underlies the performance gains still remains unclear. Intuitively, we presume that contrasting different graph augmentations can capture the essential information existing in the original user-item interactions, by randomly removing the redundancy and impurity with the edge/node dropout. Unexpectedly, a few latest works \cite{yu2021socially,zhou2021selfcf,DBLP:conf/sigir/LeeKJPY21} have reported that even extremely sparse graph augmentations (with edge dropout rate 0.9) in CL can bring desired performance gains. Such a phenomenon is quite elusive and counter-intuitive because a large dropout rate will result in a huge loss of the raw information and a highly skewed graph structure. It naturally raises a meaningful question: \textit{Do we really need graph augmentations when integrating CL with recommendation?} \par

To answer this question, we first conduct experiments with and without the graph augmentations respectively for a performance comparison. The results show that the when the graph augmentations are absent, the performance is also comparable to those with graph augmentations. We then investigate the embedding space learned by non-CL and CL-based recommendation methods. By visualizing the distributions of the representations and associating them with their performances, we find that what really matters for the recommendation performance is the CL loss, rather than the graph augmentation. Optimizing the contrastive loss InfoNCE \cite{oord2018representation} learns more evenly distributed user/item representations no matter if graph augmentations are applied, which implicitly plays a role in mitigating the popularity bias \cite{chen2020bias}. Meanwhile, despite not as effective as expected, graph augmentations are not utterly useless in the sense that the properly perturbed versions of the original graph help learn representations invariant to the disturbance factors \cite{bachman2019learning,chen2020simple}. However, generating hand-crafted graph augmentations requires constant reconstruction of the graph adjacency matrix during training, which is quite time-consuming. In addition, dropping a critical edge/node (e.g., a cut edge) may split a connected graph into a few disconnected components, at the risk of making the augmented graph and the original graph share little learnable invariance. In view of these defects, a follow-up question then arises: \textit{Are there more effective and efficient augmentation approaches?}
\par
In this paper, we give an affirmative answer to the question. On top of our finding that the uniformity of the representation distribution is the key point, we develop a graph-augmentation-free CL method in which the uniformity is more controllable. Technically, we follow the graph CL framework presented in Fig. \ref{figure:gcl}, but we discard the dropout-based graph augmentation and instead add random uniform noises to the original representations for a representation-level data augmentation. Imposing different
random noises creates variance between contrastive views, while the learnable invariance is still retained due to the controlled magnitude. Compared with the graph augmentation, the noise version directly regularizes the embedding space towards a more even distribution, which is easy-to-implement and far more efficient.
\par    
The major contributions of this paper are summarized as follows:
\begin{itemize}[leftmargin=*]
	\item We experimentally unravel why CL can boost recommendation performance and illustrate that the InfoNCE loss, rather than the graph augmentation, is the decisive factor.   
	\item We propose a simple yet effective graph-augmentation-free CL method for recommendation that can regulate the uniformity in a smooth way. It can be an ideal alternative of cumbersome graph augmentation-based CL methods. 
	\item We conduct a comprehensive experimental study on three benchmark datasets showing that the proposed method has distinct advantages over its graph augmentation-based counterparts in terms of recommendation accuracy and model training efficiency.

\end{itemize} 

\section{Investigation of Graph Contrastive Learning in Recommendation}
\subsection{Graph CL for Recommendation}
CL is often applied to recommendation with a particular set of presumed representational invariances to data augmentations \cite{zhou2020s,wujc2021self,yu2021self,xie2020contrastive}. In this paper, we revisit the most commonly used dropout-based augmentation on graphs \cite{wujc2021self,you2020graph} which assumes that the representations are invariant to partial structure perturbations. An investigation is launched into a state-of-the-art CL-based recommendation model, SGL \cite{wujc2021self}, which performs node and edge dropout to augment the original graph and adopts InfoNCE \cite{oord2018representation} for CL. Formally, the joint learning scheme in SGL is defined as:
\begin{equation}
	\label{joint}
	\mathcal{L}_{joint} = \mathcal{L}_{rec} + \lambda\mathcal{L}_{cl},
\end{equation}
which consists of two losses: recommendation loss $\mathcal{L}_{rec}$ and CL loss $\mathcal{L}_{cl}$. The InfoNCE in SGL is formulated as:
\begin{equation}
	\label{loss:cl}
	\mathcal{L}_{cl}=\sum_{i \in \mathcal{B}}-\log \frac{\exp (\mathbf{z}_{i}^{\prime\top}\mathbf{z}_{i}^{\prime \prime} / \tau)}{\sum_{j \in \mathcal{B}} \exp (\mathbf{z}_{i}^{\prime\top}\mathbf{z}_{j}^{\prime \prime} / \tau)},
\end{equation}
where $i,\ j$ are users/items in a sampled batch $\mathcal{B}$, $\mathbf{z}^{\prime}$ ($\mathbf{z}^{\prime \prime}$) are $L_{2}$ normalized $d$-dimensional node representations learned from two different dropout-based graph augmentations, and $\tau > 0$ (e.g., 0.2) is the temperature. The CL loss encourages consistency between $\mathbf{z}_{i}^{\prime}$ and $\mathbf{z}_{i}^{\prime \prime}$ which are the augmented representations of the same node $i$ and are the positive sample of each other, while minimizing the agreement between ${z}_{i}^{\prime}$ and $\mathbf{z}_{j}^{\prime \prime}$, which are the negative samples of each other. To learn the representations from the user-item graph, SGL employs a popular and effective graph encoder LightGCN \cite{he2020lightgcn} as its backbone, whose message passing process is defined as:
\begin{equation}
	\label{lightgcn}
	\mathbf{E}=\frac{1}{1+L}(\mathbf{E}^{(0)}+\tilde{\mathbf{A}}\mathbf{E}^{(0)}+...+\tilde{\mathbf{A}}^{L}\mathbf{E}^{(0)}),
\end{equation} 
where $\mathbf{E}^{(0)}\in\mathbb{R}^{|N|\times d}$ is the randomly initialized node embeddings, $|N|$ is the number of nodes,  $L$ is the number of layers, and $\tilde{\mathbf{A}}\in\mathbb{R}^{|N|\times |N|}$ is the normalized undirected adjacency matrix. By replacing $\tilde{\mathbf{A}}$ with the adjacency matrices of the corrupted graph augmentations, $\mathbf{z}^{\prime}$ ($\mathbf{z}^{\prime \prime}$) can be learned via Eq. (3). Note that, $\mathbf{z}_{i}^{\prime}=\frac{\mathbf{e}_{i}^{\prime}}{\|\mathbf{e}_{i}^{\prime}\|_{2}}$ and $\mathbf{e}_{i}^{\prime}$ is the corrupted version of $\mathbf{e}_{i}$ in $\mathbf{E}$. For conciseness, here we just abstract the core ingredients of SGL and LightGCN. More technical details can be found in the original papers \cite{wujc2021self,he2020lightgcn}. 

\begin{table}[t]
	\caption{Performance comparison of different SGL variants.}
	\small
	\label{Table:augmentation}
	\renewcommand\arraystretch{1.0}
	\begin{center}
{
	\begin{tabular}{ccc|cc}
		\toprule
		\multirow{2}{*}{\textbf{Method}}&
		\multicolumn{2}{c}{\textbf{Yelp2018}} & \multicolumn{2}{c}{\textbf{Amazon-Book}} \cr
		\cmidrule(lr){2-3}\cmidrule(lr){4-5} & \textbf{Recall@20} & \textbf{NDCG@20} & \textbf{Recall@20} & \textbf{NDCG@20}  \\ \hline	
		
		LightGCN  & 0.0639 & 0.0525 & 0.0410 & 0.0318 \\
		SGL-ND & 0.0644 & 0.0528 & 0.0440 & 0.0346  \\
		SGL-ED & \textbf{0.0675} & \textbf{0.0555} & \textbf{0.0478} & \textbf{0.0379}   \\
		SGL-RW  & 0.0667 & 0.0547  & 0.0457 & 0.0356  \\
		SGL-WA  & \underline{0.0671} & \underline{0.0550} & \underline{0.0466} & \underline{0.0373}    \\
		CL Only  & 0.0245 & 0.0190 & 0.0314 & 0.0258    \\	
		\bottomrule
		\end{tabular}}
	\end{center}
\end{table}

\subsection{Necessity of Graph Augmentation}
To demystify how CL-based recommendation methods work, we first investigate the necessity of the graph augmentation in SGL. We construct a new variant of SGL, termed \textbf{SGL-WA} (WA stands for `without augmentation'), in which the CL loss is:
\begin{equation}
	\label{loss:cl2}
	\mathcal{L}_{cl}=\sum_{i \in \mathcal{B}}-\log \frac{\exp ( 1 / \tau)}{\sum_{j \in \mathcal{B}} \exp (\mathbf{z}_{i}^{\top}\mathbf{z}_{j} / \tau)}.
\end{equation}
Because we only learn representations over the original user-item graph, then we have $\mathbf{z}_{i}^{\prime}=\mathbf{z}_{i}^{\prime\prime}=\mathbf{z}_{i}$. The experiments for the performance comparison are conducted on two benchmark datasets: \textit{Yelp2018} and \textit{Amazon-Book} \cite{he2020lightgcn,wang2019neural}. A three-layer setting is adopted and the hyperparameters are tuned according to the original paper of SGL (more experimental details in Section 4.1). The results are presented in Table \ref{Table:augmentation}. Three variants of SGL (with dropout rate 0.1) proposed in the paper are evaluated (-ND denotes node dropout, -ED is short for edge dropout, and -RW means random walk (i.e., multi-layer edge dropout). CL Only means that only the CL loss in SGL is minimized).

As can be observed, all the variants of SGL outperform LightGCN by large margins, which demonstrates the effectiveness of CL in improving recommendation performance. To our surprise, when the graph augmentation is detached, the performance gains are still so remarkable that SGL-WA even exhibits superiority over SGL-ND and SGL-RW. We conjecture that the node dropout and random walk (especially the former) are very likely to drop the key nodes and associated edges and hence break the correlated subgraphs into disconnected pieces, which highly distort the original graph. Such graph augmentations share little learnable invariance, and encouraging consistency between them probably has a negative impact. By contrast, one-time edge dropout is at a lower risk to largely disturb the semantics of the original graph, so that SGL-ED can maintain a trivial advantage over SGL-WA, which suggests the potential of a proper graph augmentation. However, considering the time-consuming reconstruction of the adjacency matrices in each epoch, we should rethink the necessity of graph augmentations and search for better alternatives. Besides, we wonder what underlies the outstanding performance of SGL-WA since no variances are provided in its CL part.

\begin{figure*}[t]	
	\centering
	\subfloat[Distribution of item representations learned from the dataset of Yelp2018.]{%
	  \includegraphics[clip,width=0.9\textwidth]{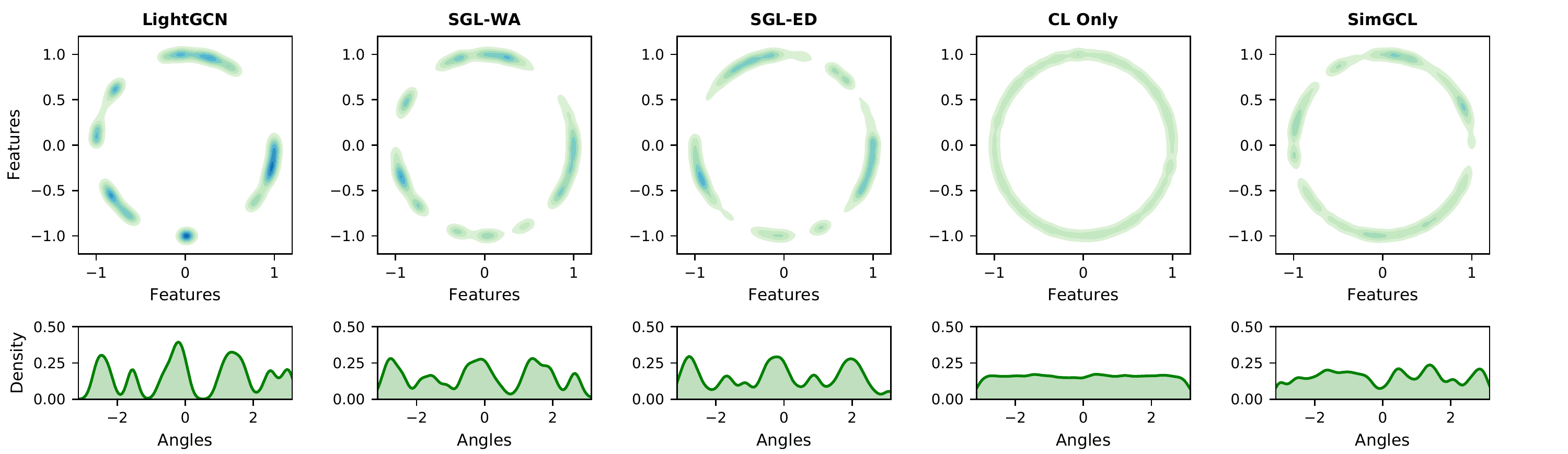}%
	}	
	\\
	\subfloat[Distribution of item representations learned from the dataset of Amazon-Book.]{%
	  \includegraphics[clip,width=0.9\textwidth]{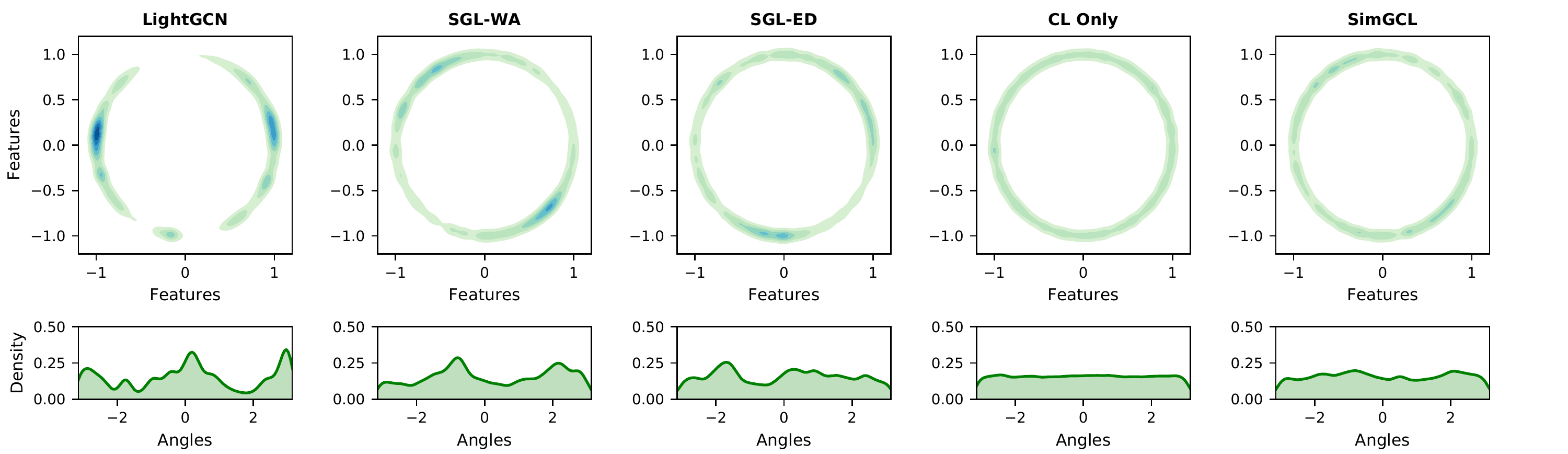}%
	}	
	\caption{We plot feature distributions with Gaussian kernel density estimation (KDE) in $\mathbb{R}^{2}$ (the darker the color is, the more points fall in that area.) and KDE on angles (i.e., arctan2(y, x) for each point (x,y) $\in \mathcal{S}^{1}$).}
	\label{figure:dist} 
\end{figure*}

\subsection{InfoNCE Loss Influences More}
Wang and Isola \cite{wang2020understanding} have identified that optimizing the contrastive loss intensifies two properties in the visual representation learning: \textit{alignment} of features from positive pairs, and \textit{uniformity} of the normalized feature distribution on the unit hypersphere. It is unclear if the CL-based recommendation methods exhibit similar patterns that can explain the results in Section 2.2. Since top-N recommendation is a one-class problem, we only investigate the uniformity by following the visualization method in \cite{wang2020understanding}. \par

We first map the learned representations (randomly sample 2,000 users for each dataset) to 2-dimensional normalized vectors on the unit hypersphere $\mathcal{S}^{1}$ (i.e., circle with radius 1) by using t-SNE \cite{van2008visualizing}. All the representations are obtained when the methods reach their best performance. Then we plot the feature distributions with the nonparametric Gaussian kernel density estimation \cite{botev2010kernel} in $\mathbb{R}^{2}$ (shown in Fig. \ref{figure:dist}). For a clearer presentation, the density estimations on angles for each point on $\mathcal{S}^{1}$ are also visualized. According to Fig. \ref{figure:dist}, we can observe notably different feature/density distributions. In the leftmost column, LightGCN shows highly clustered features that mainly reside on some narrow arcs. While in the second and the third columns, the distributions become more uniform, and the density estimation curves are less sharp, no matter if the graph augmentations are applied. In the forth column, we plot the features learned only by the contrastive loss in Eq. (2). The distributions are almost completely uniform.\par

We think that two reasons may explain the highly clustered feature distributions. The first is the message passing mechanism in LightGCN. With the increase of layers, node embeddings become locally similar. The second is the popularity bias \cite{chen2020bias} in the recommendation data. Recall the BPR loss \cite{rendle2009bpr} used in LightGCN:
\begin{equation}
	\label{bpr}
	\mathcal{L}_{rec} = -\log(\sigma(\mathbf{e}_{u}^{\top}\mathbf{e}_{i}-\mathbf{e}_{u}^{\top}\mathbf{e}_{j})),
\end{equation}
which is with a triplet input $(u,i,j)$. To optimize the BPR loss, we can get the gradients \textit{w.r.t} $\mathbf{e}_{u}$:$\nabla_{\mathbf{e}_{u}} = -\eta(1-s)(\mathbf{e}_{i}-\mathbf{e}_{j})$, where $\eta$ is the learning rate, $\sigma$ is the sigmoid function, $s = \sigma(\mathbf{e}_{u}^{\top}\mathbf{e}_{i}-\mathbf{e}_{u}^{\top}\mathbf{e}_{j})$, $\mathbf{e}_{u}$ is the user embedding, and $\mathbf{e}_{i}$ and $\mathbf{e}_{j}$ denote the positive and negative item embeddings, respectively. Since the recommendation data usually follows a long-tail distribution, when $i$ is a popular item with a large number of interactions, the user embedding will be constantly updated towards $i$'s direction (i.e., $-\nabla_{\mathbf{e}_{u}}$). The message passing mechanism further exacerbates the clustering problem (i.e., $\mathbf{e}_{u}$ and $\mathbf{e}_{i}$ aggregate information from each other in the graph convolution) and causes the representation degeneration \cite{qiu2022contrastive}.\par

As for the distributions in other columns, by rewriting Eq. (\ref{loss:cl2}), we can derive
\begin{equation}
	\mathcal{L}_{cl}=\sum_{i \in \mathcal{B}}-1 / \tau + \log\big(\exp(1 / \tau)+\sum_{j \in \mathcal{B}/\{i\}} \exp (\mathbf{z}_{i}^{\top}\mathbf{z}_{j} / \tau)\big).	
\end{equation}
Because $1/\tau$ is a constant, optimizing the CL loss is actually minimizing the cosine similarity between different nodes embeddings $\mathbf{e}_{i}$ and $\mathbf{e}_{j}$, which will push connected nodes away from the high-degree hubs in the representation space and lead to a more even distribution. \par

By associating the results in Table \ref{Table:augmentation} with the distributions in Fig. \ref{figure:dist}, we can easily draw a conclusion that the uniformity of the distribution is the underlying factor that has a decisive impact on the recommendation performance in SGL, rather than the dropout-based graph augmentations. Optimizing the CL loss can be seen as an implicit way to debias (discussed in section 4.2) because a more even representation distribution can preserve the intrinsic characteristics of nodes and improve the generalization ability. This can be a persuasive explanation for the unexpected performance of SGL-WA. It also should be noted that, by only minimizing the CL loss in Eq. (\ref{loss:cl}), a poor performance will be reached, which means that a positive correlation between the uniformity and the performance only holds in a limited scope. The excessive pursuit to the uniformity will overlook the closeness of interacted pairs and similar users/items, and impairs recommendation performance. 

\section{SimGCL: Simple Graph Contrastive Learning for Recommendation}
Based on the findings in Section 2, we speculate that by adjusting the uniformity of the learned representation in a certain scope, the optimal performance can be reached. In this section, we aim to develop a \textbf{Sim}ple \textbf{G}raph \textbf{C}ontrastive \textbf{L}earning method (\textbf{SimGCL}) for recommendation that can smoothly regulate the uniformity and provide informative variance to maximize the benefit from CL.   
\subsection{Motivation and Formulation}
Since manipulating the graph structure for a more evenly-distributed representation space is intractable and time-consuming, we shift our attention to the embedding space. Inspired by the adversarial examples \cite{goodfellow2014explaining} which are constructed by adding imperceptibly small perturbation to the input images, we directly add random noises to the representation for an efficient and effective augmentation.
\par

\begin{figure}[t]
	\centering
	\includegraphics[width=.25\textwidth]{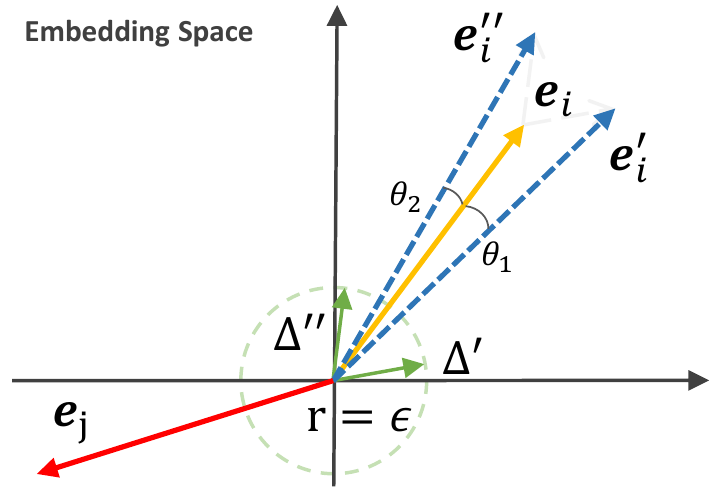}
	\caption{An illustration of the proposed random noise-based data augmentation in $\mathbb{R}^{2}$.}
	\label{figure:SimGCL}
	\vspace{-10pt}		
\end{figure}

Formally, given a node $i$ and its representation $\mathbf{e}_{i}$ in the $d$-dimensional embedding space, we can implement the following representation-level augmentation:
\begin{equation}
	\mathbf{e}_{i}^{\prime} = \mathbf{e}_{i} + \Delta^{\prime}_{i},\,\,\,\mathbf{e}_{i}^{\prime\prime} = \mathbf{e}_{i} + \Delta^{\prime\prime}_{i},
\end{equation}
where the added noise vectors $\Delta^{\prime}_{i}$ and $\Delta^{\prime\prime}_{i}$ are subject to $\|\Delta\|_{2}=\epsilon$ and $\Delta=\bar{\Delta}\odot\sign(\mathbf{e}_{i}),\,\,\bar{\Delta}\in \mathbb{R}^{d}\sim U(0,1)$. The first constraint controls the magnitude of $\Delta$, and $\Delta$ is numerically equivalent to points on a hypersphere with the radius $\epsilon$. The second constraint requires that $\mathbf{e}_{i}$, $\Delta^{\prime}$ and $\Delta^{\prime\prime}$ should be in the same hyperoctant, so that adding the noises will not cause a large deviation of $\mathbf{e}_{i}$, making less valid positive samples. In Fig. \ref{figure:SimGCL}, we illustrate Eq. (7) in $\mathbb{R}^{2}$. By adding the scaled noise vectors to the original representation, we rotate $\mathbf{e}_{i}$ by two small angles ($\theta_{1}$ and $\theta_{2}$). Each rotation corresponds to a deviation of $\mathbf{e}_{i}$, and leads to an augmented representation ($\mathbf{e}_{i}^{\prime}$ and $\mathbf{e}_{i}^{\prime\prime}$). Since the rotation is small enough, the augmented representation retains most information of the original representation and meanwhile also keeps some variance. Note that, for each node representation, the added random noises are different. 
\par
Following SGL, we adopt LightGCN as the graph encoder to propagate node information and amplify the impact of the variance due to its simple structure and effectiveness. At each layer, different scaled random noises are imposed on the current node embeddings. The final perturbed node representations are learned by:
\begin{equation}
	\begin{aligned}
		\mathbf{E}^{\prime}=\frac{1}{L}\big((\tilde{\mathbf{A}}\mathbf{E}^{(0)}+\mathbf{\Delta}^{(1)})+(\tilde{\mathbf{A}}(\tilde{\mathbf{A}}\mathbf{E}^{(0)}+\mathbf{\Delta}^{(1)})+{\Delta}^{(2)}))+...\\+(\tilde{\mathbf{A}}^{L}\mathbf{E}^{(0)}+\tilde{\mathbf{A}}^{L-1}\mathbf{\Delta}^{(1)}+...+\tilde{\mathbf{A}}\mathbf{\Delta}^{(L-1)}+\mathbf{\Delta}^{(L)})\big)
	\end{aligned}
\end{equation} 
It should be mentioned that we skip the input embedding $\mathbf{E}^{(0)}$ in all the three encoders when calculating the final representations, because we experimentally find that skipping it can lead to slight performance improvement in our setting. However, without the CL task, this operation will result in a performance drop of LightGCN. Finally, we also unify the BPR loss (Eq. (5)) and the CL loss (Eq. (2)), and then use Adam to optimize the joint loss presented in Eq. (\ref{joint}).  

\subsection{Regulating Uniformity}
In SimGCL, two hyperparameters $\lambda$ and $\epsilon$ can influence the uniformity of the representations which are critical to the performance. But $\epsilon$ can explicitly and smoothly regulate the uniformity beyond by only tuning $\lambda$. By adjusting the value of $\epsilon$, we can directly control how far the augmented representations deviate from the original. Intuitively, a larger $\epsilon$ will lead to a more roughly even distribution of the learned representation, because when the augmented representations are enough far away from the original, the information lying in their representations is also considerably influenced by the noises. As the noises are sampled from a uniform distribution, by contrasting the augmented representations, the original representation is regularized towards higher uniformity. We present the following experimental analysis to demonstrate it.
\par
In \cite{wang2020understanding}, a metric is proposed to measure the uniformity of the representation, which is the logarithm of the average pairwise Gaussian potential (a.k.a. the Radial Basis Function (RBF) kernel):
\begin{equation}
\mathcal{L}_{\text {uniform }}(f)=\log \underset{\underset{u, v\ \sim\ p_{\text {node}}}{\scriptscriptstyle i.i.d}}{\mathbb{E}}e^{-2\|f(u)-f(v)\|_{2}^{2}}.
\label{metric}
\end{equation}
where $f(u)$ outputs the $L_{2}$ normalized embedding of $u$. We choose the popular items (with more than 200 interactions) and randomly sample 5,000 users in the dataset of Yelp2018 to form the user-item pairs, and then compute the uniformity of their representations in the SGL variants and SimGCL with Eq. (\ref{metric}). For a fair comparison, a three-layer setting is applied to all the compared methods with $\lambda=0.1$. We then tune $\epsilon$ to observe how the uniformity changes. The uniformity is checked after every epoch, and we record the values in the first 30 epochs during which the compared methods all converge to their optimal solutions. \par

\begin{figure}[t]
	\centering
	\includegraphics[width=.5\textwidth]{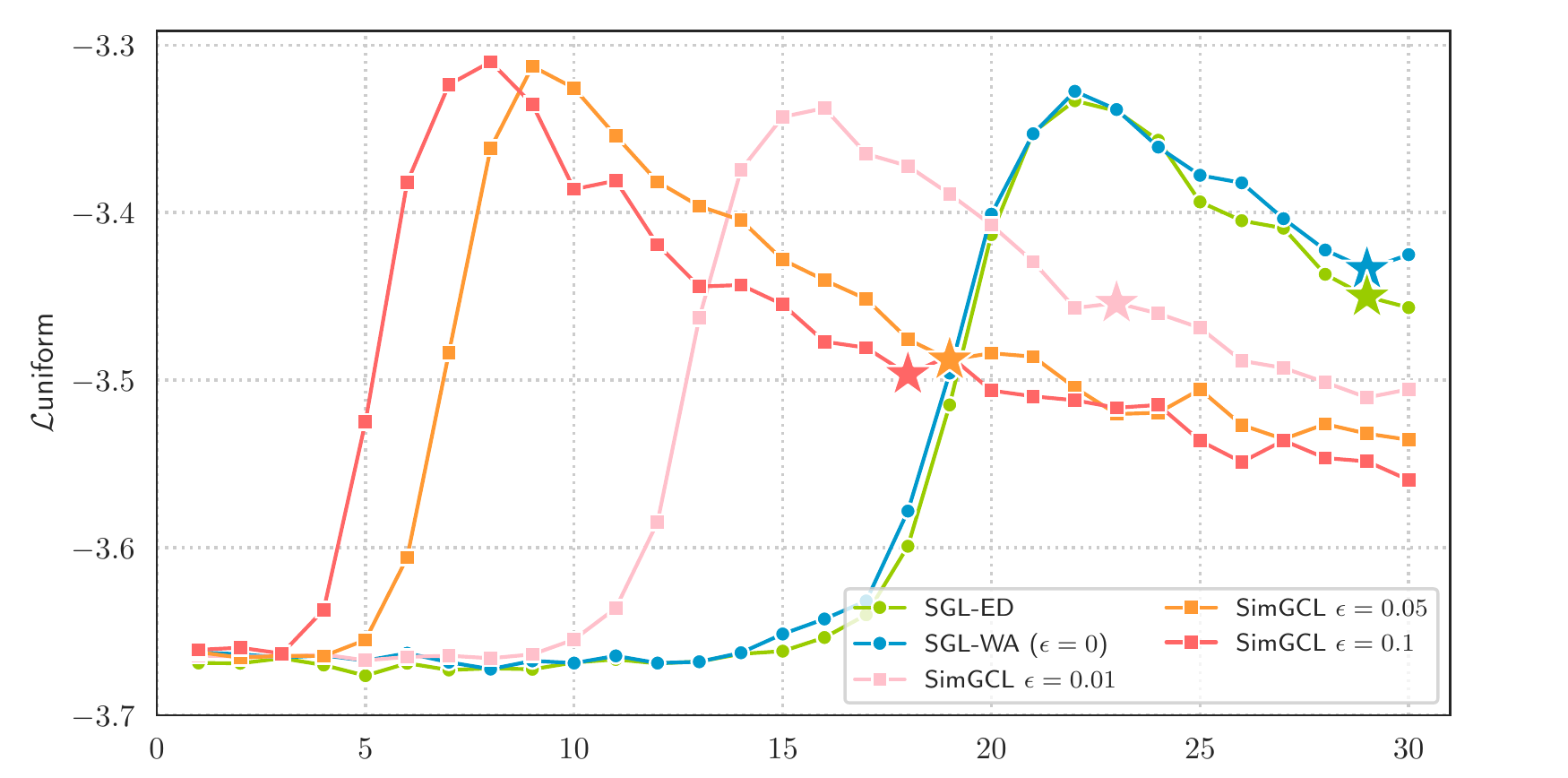}
	\caption{Trends of uniformity. The star indicates the epoch where the best recommendation performance is reached. Lower $\mathcal{L}_{\text {uniform }}$ numbers are better.}
	\label{figure:uniformity}
	\vspace{-10pt}		
\end{figure}

As clearly shown in Fig. \ref{figure:uniformity}, similar trends are observed on all the curves. At the initial stage, all the methods have highly uniformly-distributed representations because we use Xavier initialization, which is a special uniform distribution. With the training proceeding, the uniformity declines ($\mathcal{L}_{\text {uniform }}$ gets higher), and after reaching the peak, the uniformity improves till convergence and maintains this tendency. As for SimGCL, with the increase of $\epsilon$, it tends to learn more even representations, and even a very small $\epsilon=0.01$ leads to higher uniformity compared with the SGL variants. As a result, users (especially the long-tail users) are less affected by the popular items. In the rightmost column in Fig. \ref{figure:dist}, we also plot the representation distributions of SimGCL with $\epsilon=0.1$. We can clearly see that the distributions are evidently more even than those learned by SGL variants and LightGCN. All these results can support our claim that by replacing graph augmentations with the noise-based augmentations, SimGCL is more capable of controlling the uniformity of learned representations so as to debias.

\subsection{Complexity}
In this section, we analyze the time complexity of SimGCL, and compare it with that of LightGCN and its graph-augmentation based counterpart SGL-ED. We hereby discuss the batch time complexity since the in-batch negative sampling is a widely used trick in CL \cite{chen2020simple}. Let $|E|$ be the edge number in the graph, $d$ be the embedding size, $B$ denote the batch size, $M$ represent the node number in a batch, and $\rho$ denote the edge keep rate in SGL-ED. We can derive:
\begin{table}[h]
	\footnotesize
	\caption{The comparison of time complexity}
	\begin{tabular}{c|c|c|c}
	\hline	
	Component                                                                    & LightGCN          & SGL-ED            & SimGCL              \\ \hline\hline
	\makecell{Adjacency\\Matrix} & \makecell{$\mathcal{O}(2|E|)$} & \makecell{$\mathcal{O}(2|E|+4\rho|E|)$} & \makecell{$\mathcal{O}(2|E|)$} \\ \hline
    \makecell{Graph\\Convolution} & \makecell{$\mathcal{O}(2|E|Ld)$} & \makecell{$\mathcal{O}((2+4\rho)|E|Ld)$} & \makecell{$\mathcal{O}(6|E|Ld)$} \\ \hline	
	BPR Loss & $\mathcal{O}(2Bd)$ & $\mathcal{O}(2Bd)$ & $\mathcal{O}(2Bd)$ \\ \hline

    CL Loss & - & $\mathcal{O}(Bd+BMd)$ & $\mathcal{O}(Bd+BMd)$\\  \hline
	\end{tabular}
	\end{table}

\begin{table*}[t]
	\small
	\caption{Performance Comparison for different CL methods on three benchmarks.}
	\label{Table:comparison}
	\renewcommand\arraystretch{1.1}
	\begin{center}
{	
	\begin{tabular}{cc|cc|cc|cc}
		\toprule
		\multicolumn{2}{c}{\multirow{2}{*}{\textbf{Method}}}&
		\multicolumn{2}{c}{\textbf{Douban-Book}} & \multicolumn{2}{c}{\textbf{Yelp2018}} & \multicolumn{2}{c}{\textbf{Amazon-Book}} \cr
		\cmidrule(lr){3-4}\cmidrule(lr){5-6}\cmidrule(lr){7-8} && \textbf{Recall} & \textbf{NDCG} &  \textbf{Recall} & \textbf{NDCG} & \textbf{Recall} & \textbf{NDCG}  \\ \hline				
		
		\multirow{6}{*}{\textbf{1-Layer}} 
		&LightGCN &0.1394 & 0.1165 & 0.0631 & 0.0515 & 0.0384 & 0.0298 \\
		&SGL-ND &0.1619 (+16.1\%) & 0.1448 (+24.3\%) & \underline{0.0643 (+1.9\%)} & \underline{0.0529 (+2.7\%)} & 0.0432 (+12.5\%) & 0.0334 (+12.1\%)  \\
		&SGL-ED &\underline{0.1658 (+18.9\%)} & \underline{0.1491 (+28.0\%)} & 0.0637 (+1.0\%) & 0.0526 (+2.1\%)& \underline{0.0451 (+17.4\%)} & \underline{0.0353 (+18.5\%)}  \\
		&SGL-RW &\underline{0.1658 (+18.9\%)} & \underline{0.1491 (+28.0\%)} & 0.0637 (+1.0\%) & 0.0526 (+2.1\%) & \underline{0.0451 (+17.4\%)} & \underline{0.0353 (+18.5\%)}   \\
		&SGL-WA &0.1628 (+16.8\%) & 0.1454 (+24.8\%) & 0.0628 (-0.4\%) & 0.0525 (+1.9\%) & 0.0403 (+4.9\%) & 0.0320 (+7.4\%)  \\
		&\textbf{SimGCL} &\textbf{0.1720 (+23.4\%)} & \textbf{0.1519 (+30.4\%)} & \textbf{0.0689 (+9.2\%)} & \textbf{0.0572 (+11.1\%)} & \textbf{0.0453 (+18.0\%)}& \textbf{0.0358 (+20.1\%)}  \\
		\hline	
		
		\multirow{6}{*}{\textbf{2-Layer}} 
		&LightGCN &0.1485 & 0.1272 & 0.0622 & 0.0504 & 0.0411 & 0.0315 \\
		&SGL-ND &0.1622 (+9.2\%) & 0.1434 (+12.7\%)& 0.0658 (+5.8\%)& 0.0538 (+6.7\%) & 0.0427 (+3.9\%) & 0.0335 (+6.3\%)  \\
		&SGL-ED &\underline{0.1721 (+15.9\%)} & \underline{0.1525 (+19.9\%)} & \underline{0.0668 (+7.4\%)} & \underline{0.0549 (+8.9\%)} & \underline{0.0468 (+13.9\%)} & \underline{0.0371 (+17.8\%)}   \\
		&SGL-RW &0.1710 (+15.2\%) & 0.1516 (+19.2\%) & 0.0644 (+3.5\%) & 0.0530 (+5.2\%) & 0.0453 (+10.2\%) & 0.0358 (+13.7\%)   \\
		&SGL-WA &0.1687 (+13.6\%)& 0.1501 (+18.0\%) & 0.0653 (+5.0\%) & 0.0544 (+7.9\%) & 0.0453 (+10.2\%) & 0.0358 (+13.7\%)   \\
		&\textbf{SimGCL}  & \textbf{0.1770 (+19.2\%)} & \textbf{0.1582 (+24.4\%)} & \textbf{0.0719 (+15.6\%)}  & \textbf{0.0601 (+19.2\%)} & \textbf{0.0507 (+23.4\%)} & \textbf{0.0405 (+28.6\%)}   \\
		\hline	
		\multirow{6}{*}{\textbf{3-Layer}} 
		&LightGCN &0.1501 & 0.1282 & 0.0639 & 0.0525 & 0.0410 & 0.0318 \\
		&SGL-ND &0.1626 (+8.3\%) & 0.1450 (+13.1\%) & 0.0644 (+0.8\%) & 0.0528 (+0.6\%) & 0.0440 (+7.3\%)& 0.0346 (+8.8\%)  \\
		&SGL-ED &\underline{0.1732 (+15.4\%)}& \underline{0.1551 (+21.0\%)}& \underline{0.0675 (+5.6\%)} & \underline{0.0555 (+5.7\%)} & \underline{0.0478 (+16.6\%)} & \underline{0.0379 (+19.2\%)}   \\
		&SGL-RW &0.1730 (+15.3\%)& 0.1546 (+20.6\%)& 0.0667 (+4.4\%)& 0.0547 (+4.2\%)& 0.0457 (+11.5\%)& 0.0356 (+12.0\%)  \\
		&SGL-WA &0.1705 (+12.0\%)& 0.1525 (+19.0\%)& 0.0671 (+5.0\%)& 0.0550 (+4.8\%)& 0.0466 (+13.7\%)& 0.0373 (+18.4\%)  \\
		&\textbf{SimGCL} &\textbf{0.1772 (+18.1\%)} & \textbf{0.1583 (+23.5\%)} & \textbf{0.0721 (+12.8\%)} & \textbf{0.0601 (+14.5\%)} & \textbf{0.0515 (+25.6\%)} & \textbf{0.0414 (+30.2\%)}  \\	
		
		\bottomrule
		\end{tabular}}
	\end{center}
\end{table*}

	\begin{itemize}[leftmargin=*]
		\item For LightGCN and SimGCL, no graph augmentations are required, so they just need to normalize the original adjacency matrix which has $2|E|$ non-zero elements. For SGL-ED, two graph augmentations are used and each has 2$\rho|E|$ non-zero elements.
		\item In the graph convolution stage, a three-encoder architecture (see Fig. 1) is employed in both SGL-ED and SimGCL to learn augmented node representations. So, the time costs of SGL-ED and SimGCL are almost three times that of LightGCN.
		\item As for the recommendation loss, three methods all use the BPR loss and each batch contains $B$ interactions, so they have the same time cost in this component.
		\item When calculating the CL loss, the computation costs between the positive/negative samples are $\mathcal{O}(Bd)$ and $\mathcal{O}(BMd)$, respectively, because each node only considers itself as the positive, while the other nodes all are negatives.
	\end{itemize}

	Comparing SimGCL with SGL-ED, we can clearly see that SGL-ED theoretically spends less time for graph convolution, and this bonus may offset SimGCL's advantage for the adjacency matrix construction. However, when putting them into practice, we actually observe that SimGCL is much more time-efficient. That is because, the computation for the graph convolution is mostly finished on GPUs, while the graph perturbation is performed on CPUs. Besides, in each epoch, the adjacency matrices of graph augmentations in SGL-ED need to be reconstructed. While in SimGCL, the adjacency matrix of the original graph only needs to be generated once before the training. In a nutshell, SimGCL is far more efficient than SGL, beyond what we can observe from the theoretical analysis. 
	


	

\section{Experimental Results}
\subsection{Experimental Settings} 
\noindent\textbf{Datasets.} Three public benchmark datasets: Douban-Book \cite{yu2021socially} (\#user 13,024, \#item 22,347, \#interaction 792,062), Yelp2018 \cite{he2020lightgcn} (\#user 31,668 \#item 38,048, \#interaction 1,561,406), and Amazon-Book \cite{wujc2021self} (\#user 52,463, \#item 91,599, \#interaction 2,984,108) are used in our experiments to evaluate SimGCL. Because we focus on the Top-N recommendation, following the convention in the previous research \cite{yu2021self,yu2021self}, we discard ratings less than 4 in Douban-Book, which is with a 1-5 rating scale, and reset the rest to 1. We split the datasets into three parts (training set, validation set, and test set) with a ratio of 7:1:2. Two common metrics: Recall@$K$ and NDCG@$K$ are used and we set $K$=20. For a rigorous and unbiased evaluation, each experiment in this section is conducted 5 times with ranking all the items and we then report the average result. \par
\noindent\textbf{Baselines.} Besides LightGCN and the SGL variants, the following recent data augmentation-based methods are compared.
\begin{itemize}[leftmargin=*]
	\item \textbf{Mult-VAE} \cite{liang2018variational} is a variational autoencoder-based recommendation model. It can be seen as a special self-supervised recommendation model because it has a reconstruction objective.
	\item \textbf{DNN+SSL} \cite{yao2021self} is a recent DNN-based recommendation method which adopts the similar architecture in Fig. \ref{figure:gcl}, and conducts feature masking for CL. 
	\item \textbf{BUIR} \cite{DBLP:conf/sigir/LeeKJPY21} has a two-branch architecture which consists of a target network and an online network, and only uses positive examples for self-supervised learning.
	\item \textbf{MixGCL} \cite{huang2021mixgcf} designs the hop mixing technique to synthesize hard negatives for graph collaborative filtering by embedding interpolation.
\end{itemize} 
\noindent\textbf{Hyperparameters.} For a fair comparison, we refer to the best hyperparameter settings reported in the original papers of the baselines and then fine-tune all the hyperparameters of the baselines with the grid search. As for the general settings of all the baselines, the Xavier initialization is used on all the embeddings. The embedding size is 64, the parameter for $L_{2}$ regularization is 10$^{-4}$ and the batch size is 2048. We use Adam with the learning rate 0.001 to optimize all the models. In SimGCL and SGL, we empirically let the temperature $\tau=0.2$, and this value is also reported as the best in the original paper of SGL. 
\par

\subsection{SGL \textit{vs.} SimGCL: From a Comprehensive Perspective}
As one of the core claims of this paper is that graph augmentations are not indispensable and inefficient in CL-based recommendation, in this part, we conduct a comprehensive comparison between SGL and SimGCL in terms of the recommendation performance, convergence speed, running time, and ability to debias.
\subsubsection{Performance Comparison}
We first further compare SGL with SimGCL on three different datasets with different layer settings. We do not report the results with deeper layers because both SGL and SimGCL reach their best performances with shallow structures. The used hyperparameters are reported in section 4.3. We thicken the figures representing the best performance and underline the second best. The improvements are calculated by using LightGCN as the baseline. According to Table \ref{Table:comparison}, we can draw the following observations and conclusions:
\begin{itemize}[leftmargin=*]
\item All the SGL variants and SimGCL are effective in improving LightGCN under different settings. The largest improvements are observed on Amazon-Book where SimGCL can remarkably improve LightGCN by 25.6\% on Recall, and 30.2\% on NDCG with a 3-layer setting. 
\item SGL-ED is the most effective variant of SGL while SGL-ND is the least effective. When a 2-layer or 3-layer setting is used, SGL-WA outperforms SGL-ND in most cases and shows advantages over SGL-RW in a few cases. These results demonstrate that the CL loss is the main driving force of the performance improvement while intuitive graph augmentations may not be as effective as expected, and some of them may even lower the performance.
\item SimGCL shows the best performance in all the cases, which proves the effectiveness of the random noised-based data augmentation. Particularly, on the two larger datasets: Yelp2018 and Amazon-Book, SimGCL significantly outperforms the SGL variants by large margins.
\end{itemize}

\subsubsection{Convergence Speed Comparison}
In this part, we show that SimGCL converges much faster than SGL does. A 2-layer setting is used in this part and the other parameters remain unchanged. \par
\begin{figure}[t]
	\centering
	\includegraphics[width=.5\textwidth]{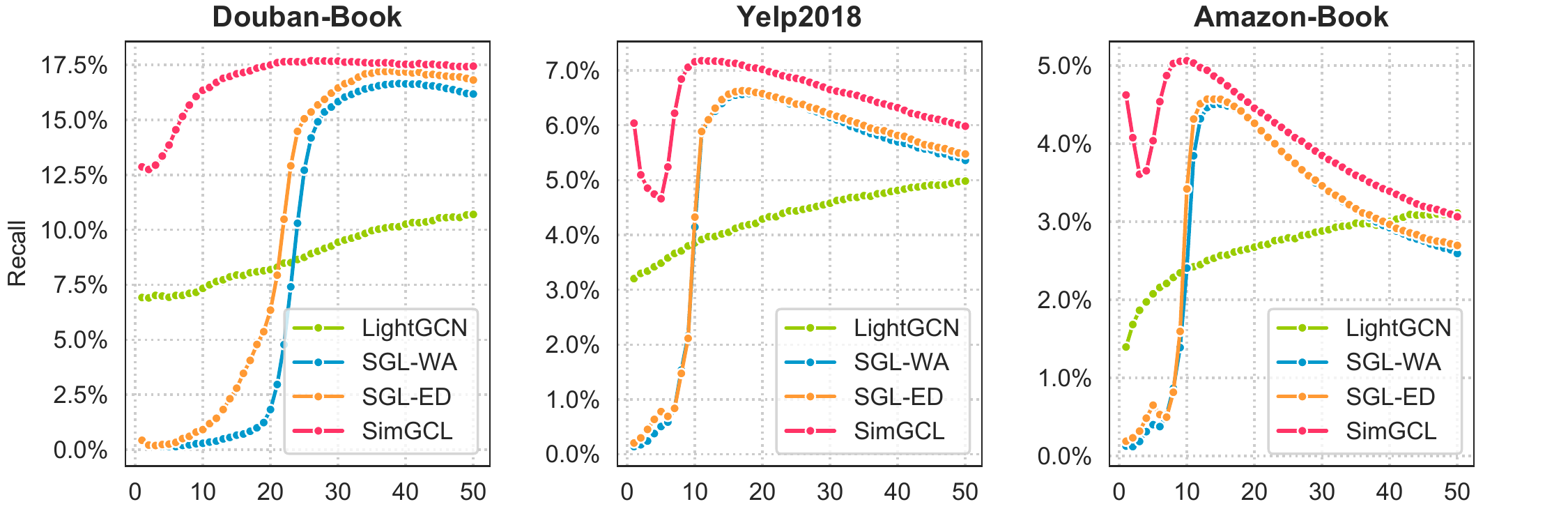}
	\caption{The performance curves in the first 50 epochs.}
	\label{figure:performance}		
\end{figure}
\begin{figure}[t]
	\centering
	\includegraphics[width=.5\textwidth]{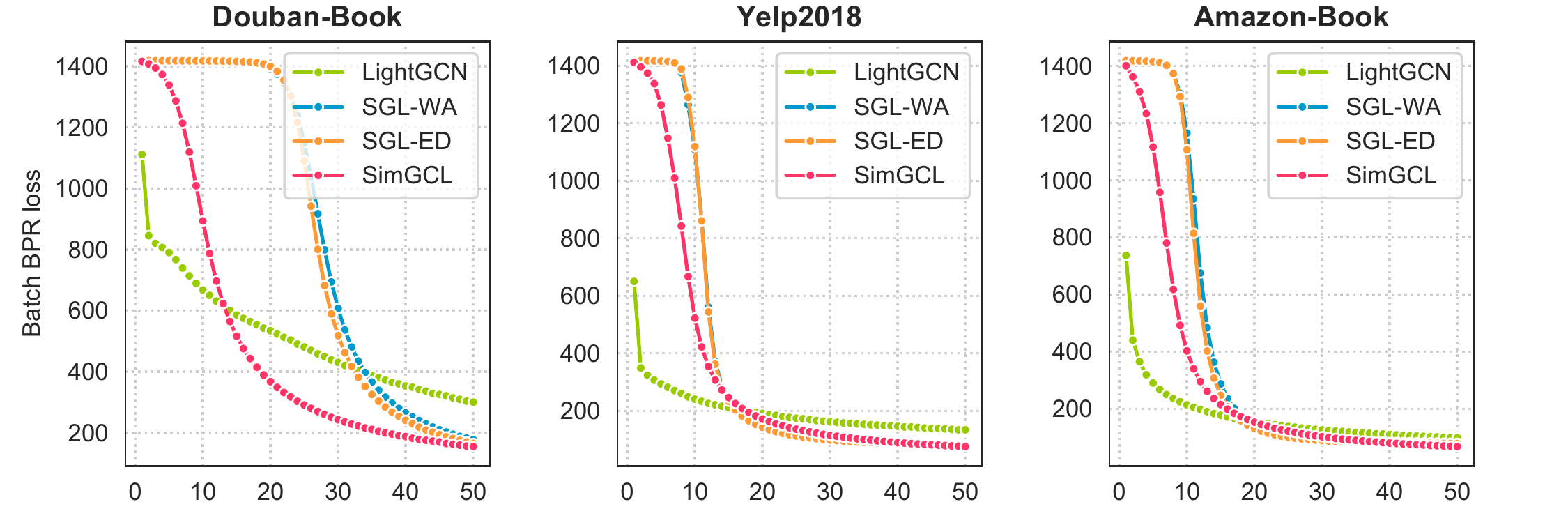}
	\caption{The loss curves in the first 50 epochs.}
	\label{figure:loss}	
\end{figure}

According to Fig. \ref{figure:performance} and Fig. \ref{figure:loss}, we can observe that, SimGCL reaches its best performance on the test set at the 25$^{\text{th}}$, the 11$^{\text{th}}$, and the 10$^{\text{th}}$ epoch on Douban-Book, Yelp2018, and Amazon-Book, respectively. By contrast, SGL-ED peaks at the 38$^{\text{th}}$, the 17$^{\text{th}}$, and the 14$^{\text{th}}$ epoch, respectively. SimGCL only spends 2/3 epochs that the SGL variants need. Besides, the curve of SGL-WA almost overlaps that of SGL-ED on Yelp2018 and Amazon-Book, and exhibits the same tendency to convergence. It seems that the dropout-based graph augmentations cannot speed up the model for a faster convergence. Despite that, all the CL-based methods show advantages over LightGCN at the convergence speed. When the other three methods begin to get overfitted, LightGCN is still hundreds of epochs distant from getting converged. \par
In the paper of SGL, the authors guess that the multiple negatives in the CL loss may contribute to the fast convergence. However, with almost infinite negative samples created by dropout, SGL-ED is basically on par with SGL-WA in speeding up the training, though the latter only has a certain number of negative samples. As for SimGCL, we consider that the remarkable convergence speed stems from the noises. By analyzing the gradients from the CL loss, we find that the noises averagely provide a constant increment, working like a momentum. In addition to the results in Fig. \ref{figure:performance} and \ref{figure:loss}, we also find that the training accelerates with $\epsilon$ getting larger. But when it is overlarge (e.g., greater than 1), despite the rapid decrease of BPR loss, SimGCL requires more time to get converged. A large $\epsilon$ acts like a large learning rate, causing the progressive \textit{zigzag} optimization that will overshoot the minimum.         

\begin{table}[h]
	\caption{Running time for per epoch (x in the brackets represents times).}
	\label{Table:time}
	\renewcommand\arraystretch{1.0}
	\begin{center}
{
	\begin{tabular}{c|c|c|c}
		\toprule
		\multirow{2}{*}{\textbf{Method}}& \textbf{Douban-Book}&\textbf{Yelp2018} & \textbf{Amazon-Book} \\ &\textbf{Time (s)} &  \textbf{Time (s)} & \textbf{Time (s)}  \\ \hline	
		
		LightGCN  &3.6 & 13.6 &  41.5 \\		
		SGL-WA & 4.4 (1.2x)& 16.3 (1.2x)& 47.0 (1.1x)  \\
		SGL-ED &13.3 (3.7x)&62.3 (4.6x)& 235.3 (5.7x)\\
		SimGCL  & 6.1 (1.7x)&27.9 (2.1x) & 98.4 (2.4x)\\			
		\bottomrule
		\end{tabular}}
	\end{center}
\end{table}

\subsubsection{Running Time Comparison}
In this part, we report the real running time that the compared methods cost for one epoch. The results in Table \ref{Table:time} are collected on an Intel(R) Xeon(R) Gold 5122 CPU and a GeForce RTX 2080Ti GPU.\par
As shown in Table \ref{Table:time}, we calculate how many times slower the other methods are when compared with LightGCN. Because there is no graph augmentation in SGL-WA, we can see its running speed is very close to that of LightGCN. For SGL-ED, two graph augmentations are required and the computation in this part is mostly finished on CPUs, so that it is even 5.7 times slower than LightGCN on Amazon-Book. The running time increases with the volume of the datasets. By contrast, despite not as fast as SGL-WA, SimGCL is only 2.4 times slower than LightGCN on Amazon-Book, and the growth trend is far lower than that of SGL-ED. Considering that SimGCL only needs 2/3 the epochs that SGL-ED spends, it outperforms SGL in all aspects \textit{w.r.t} efficiency.

\subsubsection{Ability to Debias}
\begin{figure}[t]
	\centering
	\includegraphics[width=.5\textwidth]{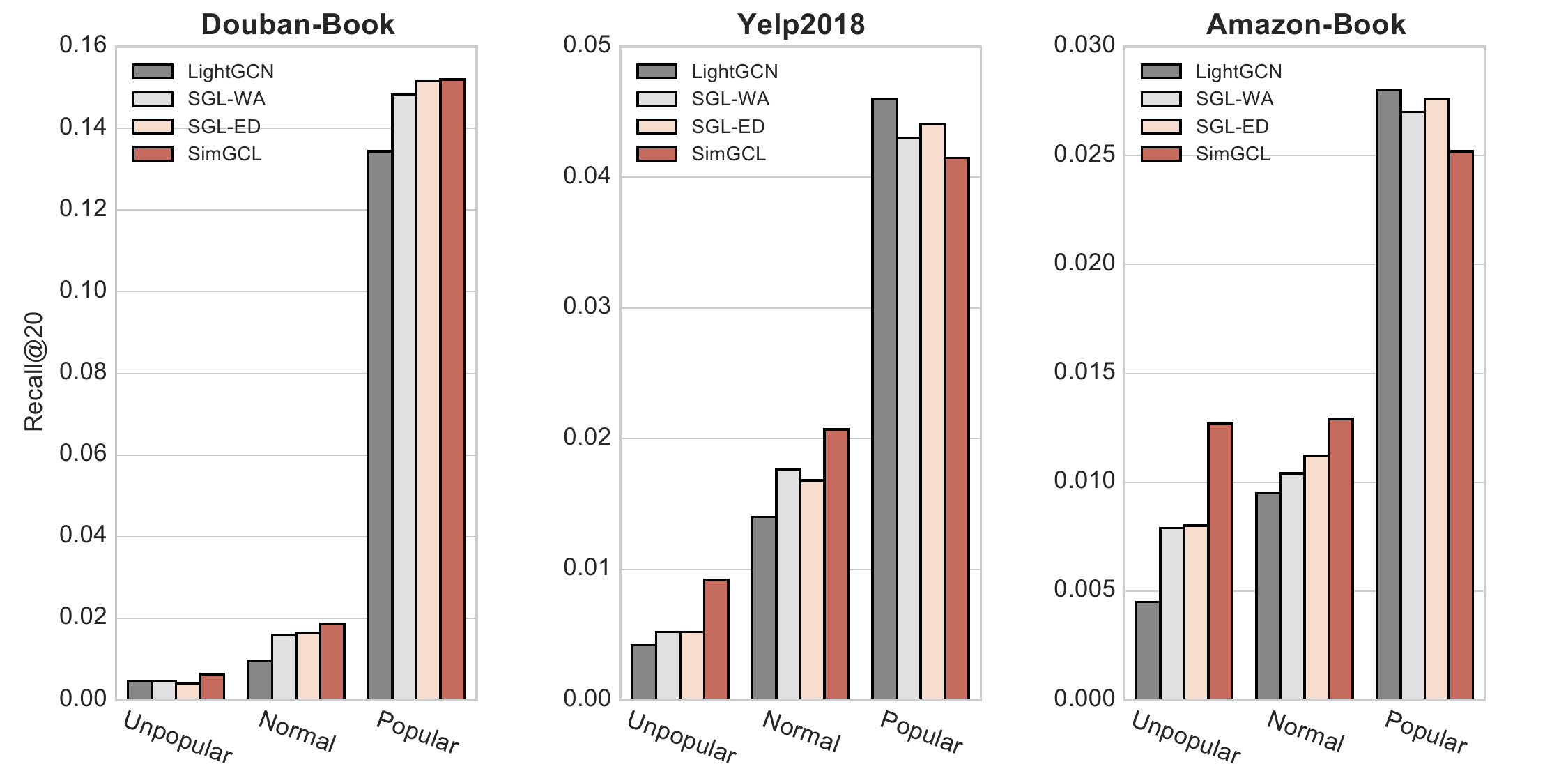}
	\caption{Performance comparison over different item groups}
	\label{figure:popularity}	
\end{figure}

The InfoNCE loss is found to have the ability to implicitly alleviate the popularity bias by learning more even representations. To verify that SimGCL upgrades this ability with the noise-based representation augmentation, we divide the test set into three subsets in proportion to the popularity of items. 80\% items with the fewest number of clicks/purchases are labelled `Unpopular', 5\% items which are most clicked/purchased are labelled `Popular', and the rest items are labelled `Normal'. We then conduct experiments to check the Recall@20 value that each group contributes (overall Recall@20 value is the sum of the values from three groups). The results are illustrated in Fig. \ref{figure:popularity}. \par
We can clearly see that the SimGCL's improvements all come from the items with lower popularity. Its prominent advantage on recommending long-tail items largely compensates for its loss on the `Popular' group. By contrast, LightGCN is inclined to recommend popular items and achieves the highest recall value on the last two datasets. The SGL variants fall between LightGCN and SimGCL on exploring long-tail items and exhibit similar recommendation preference. Combining Fig. \ref{figure:dist} with Fig. \ref{figure:popularity}, we can easily find that there is a positive correlation between the uniformity of representations and the ability to debias. Since the popular items probably have been exposed to users from other sources, recommending them may no be a good choice. On this point, SimGCL significantly outperforms SGL, and its extraordinary performance on discovering long-tail items fits the real need of users.

\subsection{Parameter Sensitivity Analysis}
In this part, we investigate the impact of the two important hyperparameters in SimGCL. Here we adopt the experimental settings used in section 4.2.2.
\subsubsection{Impact of $\lambda$} By fixing $\epsilon$ at 0.1, we change $\lambda$ to a set of predetermined representative values presented in Fig. \ref{figure:lambda}. As can be observed, with the increase of $\lambda$, the performance of SimGCL starts to increase at the beginning, and it gradually reaches its peak when $\lambda$ is 0.2 on Douban-Book, 0.5 on Yelp2018, and 2 on Amazon-Book. Afterwards, it starts to decline. Besides, in contrast to Fig. \ref{figure:eps}, more dramatic changes are observed in Fig. \ref{figure:lambda} though $\epsilon$ and $\lambda$ are tuned in the same scope, which demonstrates that $\epsilon$ can provide a finer-grained regulation beyond that provided only by tuning $\lambda$. \par

\subsubsection{Impact of $\epsilon$} 
\begin{figure}[t]
	\centering
	\includegraphics[width=.5\textwidth]{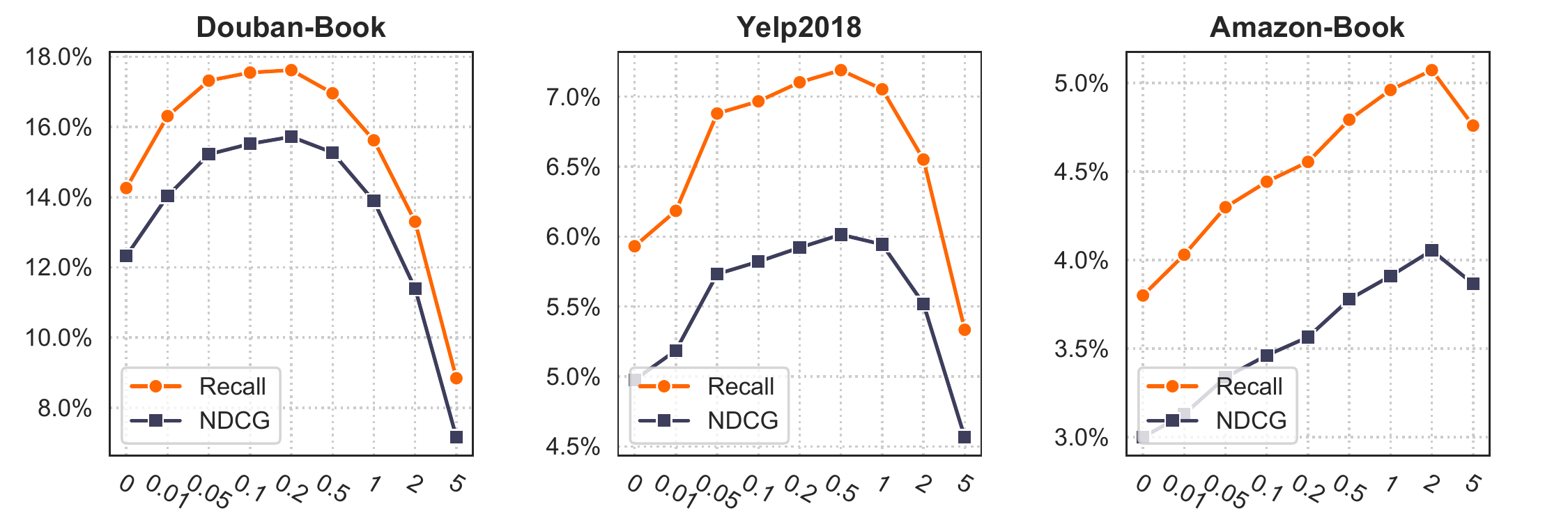}
	\caption{Influence of the magnitude $\lambda$ of CL. }
	\label{figure:lambda}
\end{figure}

\begin{figure}[h]
	\centering
	\includegraphics[width=.5\textwidth]{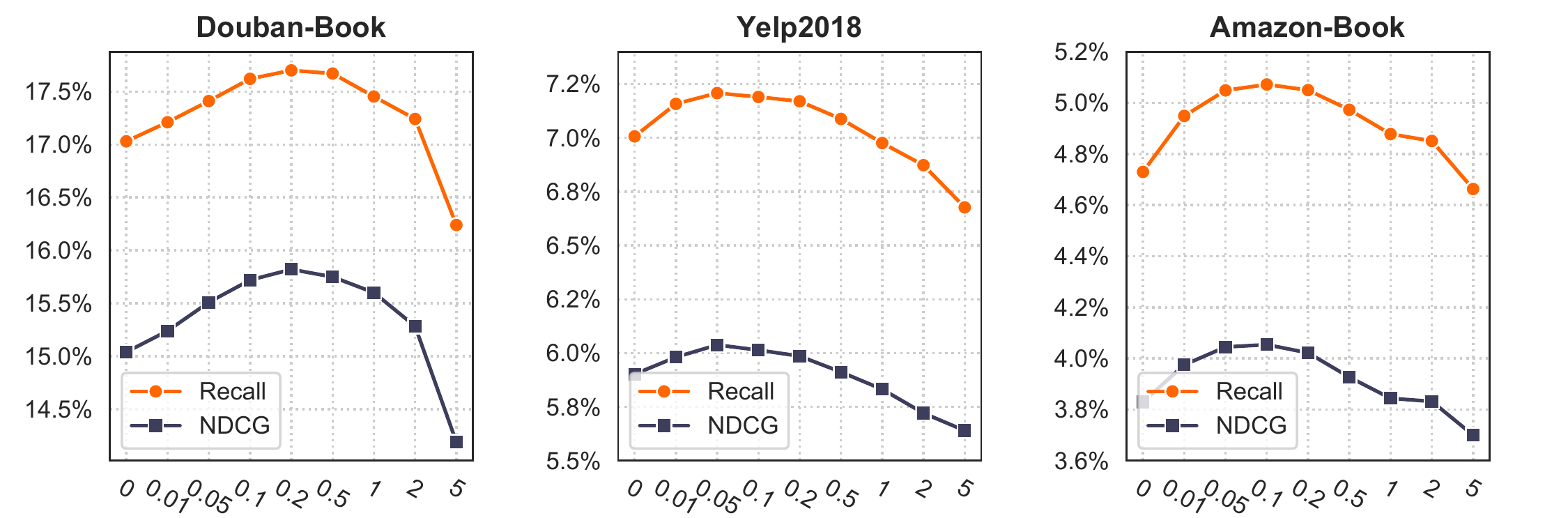}
	\caption{Influence of the noise magnitude $\epsilon$. }
	\label{figure:eps}	
	\vspace{-10pt}	
\end{figure}

We think a larger $\epsilon$ leads to a more even distribution that can help to debias. However, when it is too large, the recommendation task will be hindered because the high similarity between connected nodes cannot be reflected by an over-uniform distribution. We fix $\lambda$ at the best values on the three datasets as reported in Fig. \ref{figure:lambda}, and then adjust $\epsilon$ to see the performance change. As shown in Fig. \ref{figure:eps}, the shapes of the curves are as expected. On all the datasets, when $\epsilon$ is near 0.1, SimGCL achieves the best performance. We also find that initializing embeddings with uniform distributions (including Xavier initialization) leads to 3\%-4\% performance improvement compared with Gaussian initialization.

\subsection{Performance Comparison with Other Methods}
To further confirm the outstanding competence of SimGCL, we compare it with other four recently proposed data augmentation-based methods. According to Table \ref{Table:sota}, SimGCL outperforms all the baselines, and MixGCF is the runner-up. Meanwhile, we find that some data augmentation-based recommendation methods are not as powerful as expected, and even outperformed by LightGCN in many cases. We attribute their failure to: (1). LightGCN, MixGCF, and SimGCL are all based on the graph convolution mechanism, which are more capable of modeling graph data compared with Mult-VAE. (2). DNNs are proved effective when user/item features are available. In our datasets, no features are provided and we mask embeddings learned by DNN to conduct self-supervised learning, so it cannot fulfill itself in this situation. (3). In the paper of BUIR, it removes long-tail nodes to achieve a good performance, but we use all the users and items. Besides, its siamese network structure may also collapse to a trivial solution on some long-tail nodes because it does not use negative examples, which may account for its incompetence.   

\begin{table}[h]
	\caption{Performance comparison with other SOTA models.}
	\small
	\label{Table:sota}
	\renewcommand\arraystretch{1.0}
	\begin{center}
{
	\begin{tabular}{c|cc|cc|cc}
		\toprule
		\multirow{2}{*}{\textbf{Method}}&\multicolumn{2}{c}{\textbf{Douban-Book}}& \multicolumn{2}{c}{\textbf{Yelp2018}} & \multicolumn{2}{c}{\textbf{Amazon-Book}} \cr
		\cmidrule(lr){2-3}\cmidrule(lr){4-5}\cmidrule(lr){6-7}&\textbf{Recall} & \textbf{NDCG}  & \textbf{Recall} & \textbf{NDCG} & \textbf{Recal} & \textbf{NDCG}  \\ \hline
		LightGCN  &0.1501&0.1282& 0.0639 & 0.0525 & 0.0411 & 0.0315	\\
		Mult-VAE  &0.1310&0.1103& 0.0584 & 0.0450 & 0.0407 & 0.0315 \\	
		DNN+SSL  &0.1366&0.1148& 0.0483  & 0.0382  & 0.0438 & 0.0337  \\
		BUIR  &0.1127&0.8938& 0.0487  & 0.0404  & 0.0260 & 0.0209  \\	
		MixGCF  &\underline{0.1731}&\underline{0.1552}& \underline{0.0713}  & \underline{0.0589}  & \underline{0.0485} &\underline{0.0378}  \\		
		SimGCL &\textbf{0.1772} & \textbf{0.1583}& \textbf{0.0721} & \textbf{0.0601} & \textbf{0.0515} & \textbf{0.0410}   \\			
		\bottomrule
		\end{tabular}}
	\end{center}
\end{table}

\subsection{Performance Comparison with Different Types of Noises}
In SimGCL, we use the random noises sampled from a uniform distribution to implement data augmentation. However, there are other types of noises including the Gaussian noises and adversarial noises. Here we also test different noises and report their best results in Table \ref{Table:noise} ($_{u}$ denotes uniform noises sampled from $U(0,1)$, $_{p}$ represents the positive uniform noises which differ from $_{u}$ in not satisfying the second constraint in section 3.1, $_{g}$ denotes Gaussian noises generated by the standard Gaussian distribution, and $_{a}$ denotes adversarial noises generated by FGSM \cite{goodfellow2014explaining}). According to Table \ref{Table:noise}, SimGCL$_{_g}$ shows comparable performance while SimGCL$_{_a}$ is less effective. The possible reason is that we apply $L_{2}$ normalization to the noises. The normalized noises generated by the standard Gaussian distribution can fit a much flatter Gaussian distribution (can be easily proved) which approximates a uniform distribution. So, the comparable results are observed. As for SimGCL$_{_a}$, the adversarial noises are generated by only targeting maximizing the CL loss while the recommendation loss has a dominant status that impacts the performance more during optimization. As for SimGCL$_{_p}$, we notice a slight performance drop compared with SimGCL$_{_u}$ in most cases, which suggests the necessity of the directional constraint for creating more informative augmentations.   
\begin{table}[h]
	\small
	\caption{Performance comparison between different SimGCL variants.}
	\label{Table:noise}
	\renewcommand\arraystretch{1.0}
	\begin{center}
{
	\begin{tabular}{c|cc|cc|cc}
		\toprule
		\multirow{2}{*}{\textbf{Method}}&\multicolumn{2}{c}{\textbf{Douban-Book}}& \multicolumn{2}{c}{\textbf{Yelp2018}} & \multicolumn{2}{c}{\textbf{Amazon-Book}} \cr
		\cmidrule(lr){2-3}\cmidrule(lr){4-5}\cmidrule(lr){6-7}&\textbf{Recall} & \textbf{NDCG}  & \textbf{Recall} & \textbf{NDCG} & \textbf{Recal} & \textbf{NDCG}  \\ \hline
		LightGCN  &0.1485&0.1272& 0.0639& 0.0525 & 0.0411 & 0.0315	\\
		SimGCL$_{_a}$ &0.1561&0.1379& 0.0604 & 0.0505 & 0.0455 & 0.0358   \\
		SimGCL$_{_p}$ &0.1751 & 0.1565& 0.0708 & 0.0593 & \underline{0.0514} & \underline{0.0409}   \\
		SimGCL$_{_g}$ &\textbf{0.1773}&\textbf{0.1586}& \underline{0.0718} & \underline{0.0599} & 0.0511 & 0.0408   \\
		SimGCL$_{_u}$ &\underline{0.1772} & \underline{0.1583}& \textbf{0.0721} & \textbf{0.0601} & \textbf{0.0515} & \textbf{0.0414}   \\		
		\bottomrule
		\end{tabular}}
	\end{center}
\end{table}
\section{Related Work}
\subsection{Graph Neural Recommendation Models}
\textit{Graph Neural Networks} (GNNs) \cite{wu2020comprehensive,gao2021graph} now have become widely acknowledged powerful architectures for modeling recommendation data. This new neural network paradigm ends the regime of MLP-based recommendation models in the academia, and boosts the neural recommender systems to a new level. A large number of recommendation models, which adopt GNNs as their bases, claim that they have achieved state-of-the-art performance \cite{yu2021self,he2020lightgcn,yu2020enhance} in different subfields. Particularly, GCN \cite{kipf2016semi}, as the most prevalent variant of GNNs, further fuels the development of the graph neural recommendation models like GCMC \cite{berg2017graph}, NGCF \cite{wang2019neural}, LightGCN \cite{he2020lightgcn}, and LCF \cite{yu2020graph}. Despite the different implementations in details, these GCN-driven models share a common idea that is to acquire the information from the neighbors in the user-item graph layer by layer to refine the target node's embeddings and fulfill graph reasoning \cite{wu2020graph}. Among these methods, LightGCN is the most popular one due to its simple structure and decent performance. Following \cite{wu2019simplifying}, it removes the redundant operations including transformation matrices and nonlinear activation functions. Such a design is proved efficient and effective, and inspires a lot of follow-up CL-based recommendation models like SGL \cite{wujc2021self} and MHCN \cite{yu2021socially}.

\subsection{Contrastive Learning in Recommendation}
As CL works in a self-supervised manner \cite{yu2022survey}, it is inherently a possible solution to the data sparsity issue \cite{yu2018adaptive,yu2019generating} in recommender systems. Inspired by the achievements of CL in other fields, there has been a wave of new research that integrates CL with recommendation \cite{zhou2020s,wujc2021self,xia2020self,yu2021socially,yu2021self,ma2020disentangled,qiu2021memory}. Zhou \textit{et al.} \cite{zhou2020s} adopted random masking on attributes and items to create sequence augmentations for sequential model pretraining with mutual information maximization. Wei \textit{et al.} \cite{wei2021contrastive} reformulated the cold-start item representation learning from an information-theoretic standpoint and maximized the mutual dependencies between item content and collaborative signals to alleviate the data sparsity issue. Similar ideas are also found in \cite{yao2021self}, where a two-tower DNN architecture is developed for recommendation, in which the item tower is also shared for contrasting augmented item features. SEPT \cite{yu2021socially} and COTREC \cite{xia2021self} further propose to mine multiple positive samples with semi-supervised learning on the perturbed graph for social/session-based recommendation. In addition to the dropout, CL4Rec \cite{xie2020contrastive} proposes to reorder and crop item segments for sequential data augmentation. Yu \textit{et al.} \cite{yu2021self}, Zhang \textit{et al.} \cite{zhang2021double} and Xia \textit{et al.} \cite{xia2020self} leveraged hypergraph to model recommendation data, and proposed to contrast different hypergraph structures for representation regularization. In addition to the data sparsity problem, Zhou \textit{et al.} \cite{zhou2021contrastive} theoretically proved that CL can also mitigate the exposure bias in recommendation, and developed a method named CLRec to improve deep match in terms of fairness and efficiency. 

\section{Conclusion}
In this paper, we revisit the dropout-based CL in recommendation, and investigate how it improves recommendation performance. We reveal that, in CL-based recommendation models, the CL loss is the core and the graph augmentation only plays a secondary role. Optimizing the CL loss leads to a more even representation distribution, which helps to debias in the scenario of recommendation. We then develop a simple graph-augmentation-free CL method to regulate the uniformity of the representation distribution in a more straightforward way. By adding directed random noises to the representation for different data augmentations and contrast, the proposed method can significantly enhance recommendation. The extensive experiments demonstrate that the proposed method outperforms its graph augmentation-based counterparts and meanwhile the training time is dramatically reduced.\par

\section*{Acknowledgement}
This work is supported by Australian Research Council Future Fellowship (Grant No. FT210100624), Discovery Project (Grant No. DP190101985) and Discovery Early Career Research Award (Grant No. DE200101465).
\bibliographystyle{ACM-Reference-Format}
\bibliography{refs}

\end{document}